\documentclass{jps-cp}
\usepackage{txfonts} 
\RequirePackage{graphicx}
\graphicspath{{figs/}}
\usepackage{wrapfig}

\title{Study of Thin Iron Films for Polarization Analysis of Ultracold Neutrons
}

\author{Hiroaki \textsc{Akatsuka}$^{1}$,  
Takashi \textsc{Higuchi}$^{2}$, 
Sean \textsc{Hansen-Romu}$^{3,4}$, 
Kichiji \textsc{Hatanaka}$^{2}$, 
Tomohiro \textsc{Hayamizu}$^{5}$, 
Masahiro \textsc{Hino}$^{6}$, 
Go \textsc{Ichikawa}$^{7}$,  
Sohei \textsc{Imajo}$^{2}$,  
Blair \textsc{Jamieson}$^{3}$, 
Shinsuke \textsc{Kawasaki}$^{7}$, 
Masaaki \textsc{Kitaguchi}$^{1}$,
Ryohei \textsc{Matsumiya}$^{2,8}$, 
Kenji \textsc{Mishima}$^{7}$  
}

\inst{$^{1}$Nagoya University, Furo-cho, Chikusa-ku, Nagoya, Aichi 464-8601, Japan\\
$^{2}$RCNP, Osaka University, 10-1 Mihogaoka, Ibaraki, Osaka 567-0047, Japan \\
$^{3}$University of Winnipeg, Winnipeg, MB R3B 2E9, Canada \\
$^{4}$University of Manitoba, Winnipeg, MB R3T 2N2, Canada \\
$^{5}$RIKEN, Wako, Saitama 351-0198, Japan \\
$^{6}$Institute for Integrated Radiation and Nuclear Science, Kyoto University, Kumatori-cho,  Osaka 590-0494, Japan\\
$^{7}$High Energy Accelerator Research Organization (KEK), Tsukuba, Ibaraki 305-0801, Japan \\
$^{8}$TRIUMF, 4004 Wesbrook Mall, Vancouver, BC V6T 2A3, Canada \\
}

\email{akatsuka@phi.phys.nagoya-u.ac.jp
}

\recdate{March 15, 2022}

\abst{
The TUCAN (TRIUMF Ultra-Cold Advanced Neutron) collaboration aims to search for the neutron electric dipole moment (nEDM) with unprecedented precision. One of the essential elements for the nEDM measurement is a polarization analyzer of ultracold neutrons (UCNs), whose main component is a magnetized thin iron film. Several thin iron films were deposited on aluminum and silicon  substrates and were characterized by vibrating sample magnetometry and cold-neutron reflectometry. 
A magnetic field required to saturate the iron film is $\sim$12$\,\mathrm{kA/m}$ for those on the aluminum  substrates and  6.4$\, \mathrm{kA/m}$ for the silicon substrates. The magnetic potential of the iron films on the Si substrate was estimated to be 2$\,\mathrm{T}$ by the neutron reflectometry, which is sufficient performance for an UCN polarization analyzer of the nEDM measurement.
}
\kword{ultracold neutrons, electric dipole moment, polarization, ion-beam sputtering, cold-neutron reflectometry}
\vspace{-3em}
\begin{document}
\maketitle
\section{Introduction}
\subsection{Background}

Searches of CP violation occupy an important place in today's particle physics. According to the  Sakharov's scenario \cite{CP}, sources of CP violation beyond the Standard Model are required to explain the baryon asymmetry of the universe. The measurement of the neutron electric dipole moment (nEDM) has great significance in this context. A finite value of the nEDM would violate time-reversal symmetry and therefore CP symmetry. Decades of experimental searches of the nEDM have set stringent upper limits on the nEDM and put severe constraints on theories beyond the Standard Model.
The current best experimental upper limit on the nEDM is 1.8$\times10^{-26}\,\mathrm {e\cdot cm} $, obtained at the Paul Scherrer Institute using ultracold neutrons (UCNs) \cite{PSI}.
UCNs refer to neutrons with kinetic energies of $\lesssim$300$\,\mathrm{neV}$ that can be stored in cells made of appropriate materials and surfaces for times on the order of 100$\,\mathrm{s}$. The limitation of the latest measurement comes from the statistical uncertainty due to the number of UCNs available for the measurement. In this context, the TUCAN (TRIUMF Ultra-Cold Advanced Neutron) collaboration is developing a high-intensity UCN source and aiming at a nEDM measurement with a sensitivity of $10^{-27}\,\,\mathrm {e\cdot cm}$.

Measurements of the nEDM are performed in an UCN storage cell under  a weak ($\sim$1\,$\mu$T) magnetic field and a strong ($\sim$10$\,\mathrm{kV/cm}$) electric field. The magnitude of the nEDM is evaluated by the difference of precession frequencies under two configurations of the electric and the magnetic fields, where they are parallel or anti-parallel.
The Ramsey's method of separated oscillating fields\cite{Ramsey} is used to measure the precession frequencies.
Here, a sequence of two $\pi/2$ pulses is applied to the UCNs and their polarization is measured at the end of every cycle. 
By observing a spin resonance by varying the frequency of the $\pi/2$ pulses, the precession frequency is determined from the center of the resonance.
Thus, the polarization analysis of UCNs is one of the essential elements of the experiment.

\subsection{Principle of UCN Polarization}
The polarization analysis of UCNs is based on a spin-dependent potential due to the magnetic moment of the neutron $\boldsymbol{\mu_n} = -$60$\,\mathrm{neV}/\mathrm{T}$ and an external magnetic flux density  ${\boldsymbol B}$. 
Magnetized thin iron films are commonly used to produce a strong magnetic potential for UCNs.  
An UCN traversing a thin iron film magnetized to a flux density of $B$ experiences a spin-dependent effective potential 
$V_{\mathrm {eff}\,\pm}$
\begin{equation}
V_{\mathrm {eff}\,\pm}=V_{\mathrm {Fe}}\pm{|\boldsymbol{\mu_n}|}\cdot {|\boldsymbol{B}|}\approx 209\,\mathrm {neV} \pm 60\,\mathrm {neV}/\mathrm{T} \cdot{|\boldsymbol{B}|}
\label{Eq:Vap}
\end{equation}
where the spin states $+$ and $-$ of the neutron are defined by the sign of the magnetic potential.
  The ideal average Fermi potential of an iron film is represented by  $V_{\mathrm {Fe}}\approx$209$\,\mathrm {neV}$.
Because the kinetic energies of UCNs are $\lesssim 300$\,neV, 
 only the $-$ state
 neutrons pass through the film.

 \subsection{Requirements for thin iron films for UCN polarization analysis}\label{sec:requirements}
 \begin{wrapfigure}{r}[0pt]{0.52\textwidth}
 \vspace{0em}
 \centering
 \includegraphics[width=0.45\textwidth]{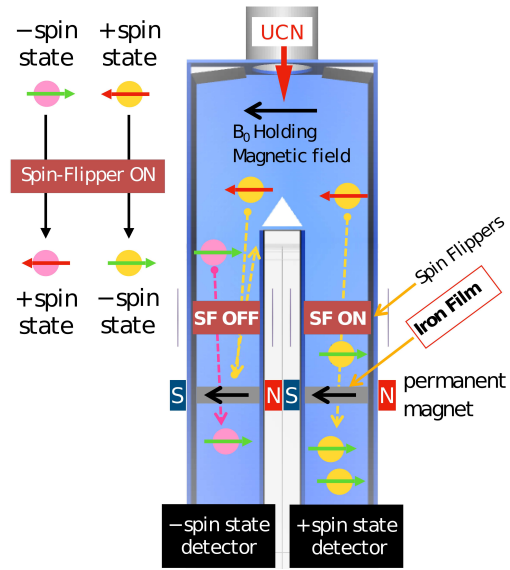}
 \caption{Schematical view of the SSA. UCNs coming into the SSA from the top are selected by their spin states and detected separately in the two arms.
 The spin flipper and the iron film in front of each detector are used to select the spin state of the neutron detected in each arm.}
 \label{SSA}
 \vspace{-1em}
\end{wrapfigure}
 In the TUCAN nEDM measurement, spin-state analysis of UCNs will be performed by a device called the simultaneous spin analyzer (SSA), which allows for simultaneous analysis of the two spin components \cite{SSA}. As shown in Fig. \ref{SSA}, it is equipped with two arms, each consisting of a spin flipper, a thin iron film with a permanent magnet surrounding it, and an UCN detector. The polarized UCNs detected by each arm can be controlled by the spin-flipper. 
 To be used for the SSA, the thin iron films are required to reach a large saturation magnetization with a smaller applied field. In the present study, we aimed to develop films that are saturated by a magnetic field smaller than 40\,kA/m of the previous studies \cite{SSA,PSI2012}. Such a small operational magnetic field will  minimize the influence of the SSA on the neighboring devices and provide flexibility in the design of the apparatus. The goal for the saturation flux density is $\sim 2$~T, which corresponds to an acceptance range of the UCN energies from 90 to 330 neV from Eq. (\ref{Eq:Vap}), sufficiently covering expected energies of UCNs detected by the SSA. 
\newpage
\label{sec:production}
 \begin{wrapfigure}{r}[0pt]{0.5\textwidth}
 \centering
 \vspace{-2em}
    \includegraphics[width=0.5\textwidth]{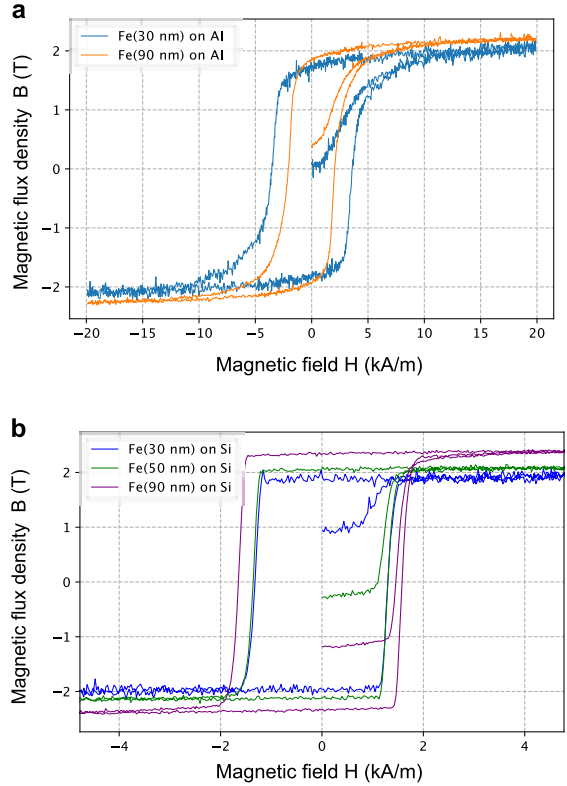}
    \caption{Magnetization curves of samples with thicknesses of 30 and 90$\,\mathrm{nm}$ on Al substrates (a) and samples with thicknesses of about 30, 50 and 90$\,\mathrm{nm}$ on Si substrates (b). 
    The magnetic moments measured by VSM are converted to the magnetic flux density by the area of the samples and the iron layer thickness estimated from the sputtering duration.
    }\label{VSM}
    \vspace{-1em}
    \end{wrapfigure}
 \section{
Production of thin iron films and evaluation of their magnetic properties
 }
 Thin iron films were simultaneously deposited on aluminum (Al) and silicon (Si) substrates by the ion beam sputtering (IBS) facility of the Institute for Integrated Radiation and Nuclear Science, Kyoto University (KURNS) \cite{Hino}. The thicknesses of the iron layers were 30, 50 and 90$\,\mathrm{nm}$. In the previous studies, thin iron films with thicknesses of 150--400$\,\mathrm{nm}$ sputtered on Al substrates were developed \cite{Lauer,SSA,PSI2012}. These thicknesses of $\lesssim$100$\,\mathrm{nm}$ were selected, as they were expected to make the iron coercivity small \cite{Kim}. In addition to the Al substrate used in the previous studies, polished Si wafers  with a diameter of 3 inches were employed as the substrates, because it is expected to have a smaller magnetization induced by the strains of the film due to the  Villari effect \cite{Villari}.
 Another advantage of  the Si substrate films is that they can be characterized by cold-neutron reflectometry, which enables direct determination of the effective potential Eq. (\ref{Eq:Vap}), as discussed in detail in Sec. \ref{sec:MLF}.
 
 The magnetic properties of the samples were measured by vibrating sample magnetometry (VSM). For the measurement, a piece of about 8$\,\mathrm{mm}\times$10$\,\mathrm{mm}$ was cut out from each sample and a magnetization curve was obtained. The obtained curves are shown in Fig. \ref{VSM}.
 It can be seen that the iron is fully saturated by a magnetic field of about 12$\,\mathrm{kA/m}$ for samples on the Al substrates and about 4$\,\mathrm{kA/m}$ for those on the  Si substrates.
 
\section{Characterization of thin iron films on Si substrate by cold-neutron reflectometry}\label{sec:MLF}
\begin{figure}[tbh]
 \centering
 \includegraphics[width=\textwidth]{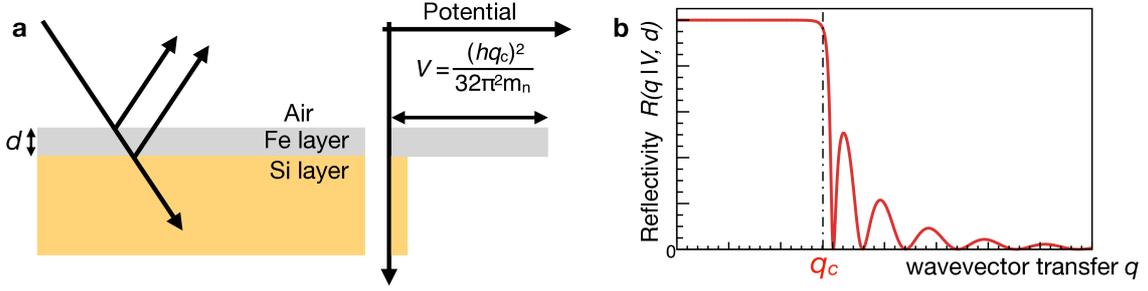}
 \centering
 \caption{\textbf{a} Reflection model of a two-layer film. \textbf{b} Typical cold-neutron reflectivity from a thin iron film on Si substrate. 
 Since the peak to peak of $R(q~|V,d)$ is $\sim \pi/d$, the thickness of the iron film $d$ can be extracted by fitting the date to the model.
 }
 \label{fig:general_R}
 \vspace{-2em}
 \end{figure}

The thin iron film samples on Si substrates with thicknesses of 
30, 50 and
90$\,\mathrm{nm}$ (20$\,\mathrm{mm} \times\,$30$\,\mathrm{mm}$) were also characterized by polarized cold-neutron reflectometry, where  the reflectivity of the sample is obtained as a function of the wavevector transfer $q$ of the neutron beam.
As described in Fig. \ref{fig:general_R}, the $q$ dependence of the reflectivity of a single-layer thin film on a sufficiently thick substrate can be described by a simple model which is characterized by the potential $V$, the thickness of the thin layer $d$, and the cut-off wavevector $q_c$.
By fitting the model $R^{\mathrm {(s)}}(q | V, d)$ to the obtained data, the effective potential $V_\mathrm{eff, \pm}$ experienced by neutrons at each spin states can be estimated.

The measurement was carried out at the low-divergence branch of J-PARC/MLF BL05 \cite{BL05}, where pulsed cold neutrons in an energy range of $0.2$--$20\,\mathrm{meV}$ are provided. The pulse structure of the beam allows for determination of the wavelength of the detected neutrons from their time of flight.
\subsection{Measurement of the beam polarization}\label{sec:polarization}
\begin{figure}[tbh]
 \centering
\includegraphics[width=0.8\textwidth]{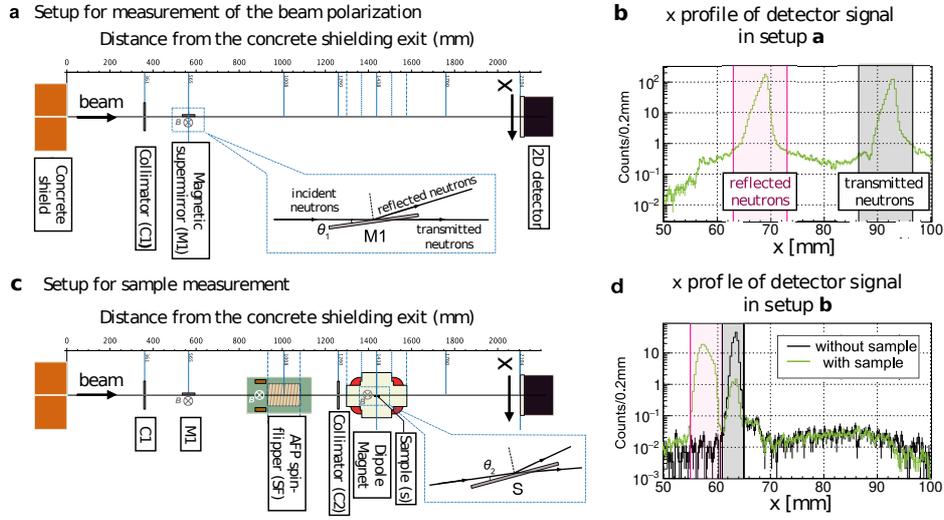}
 \caption{
  \textbf{a} Cold-neutron reflectometry setup to characterize beam polarization of the magnetic super-mirror M1. The incident angle of the M1 is $\theta_1=8.3\,\mathrm{mrad}$.
  \textbf{b} Position dependence of the neutron counts  detected by the RPMT with setup of Fig. \ref{fig:setup}{a}. The peaks corresponding to the transmitted and reflected neutrons are observed. The integration ranges to obtain the intensities of the transmitted ($I^{\mathrm{(M1)}}_T$) and reflected ($I^{\mathrm{(M1)}}_R$) neutrons are indicated by the bands in the graph.  
  \textbf{c} Setup for cold-neutron reflectometry of the samples. Incident angles of the M1 and sample are $\theta_1=10\,\mathrm {mrad}$,\,$\theta_2=12\,\mathrm {mrad}$.
  \textbf{d} Position dependence of the neutron count  in the setup Fig. \ref{fig:setup}{c}. The signals of the direct beam without the sample are shown  by the black curve. The green curve shows the results when the sample was in place. The two bands in the graph indicated the integration ranges to obtain the intensities of the direct beam  $I^{\mathrm {(s)}}_D$ and the reflected beam $I^{\mathrm {(s)}}_R$.  }
 \label{fig:setup}
\end{figure}

\begin{figure}[tbh]
 \includegraphics[width=0.9\textwidth]{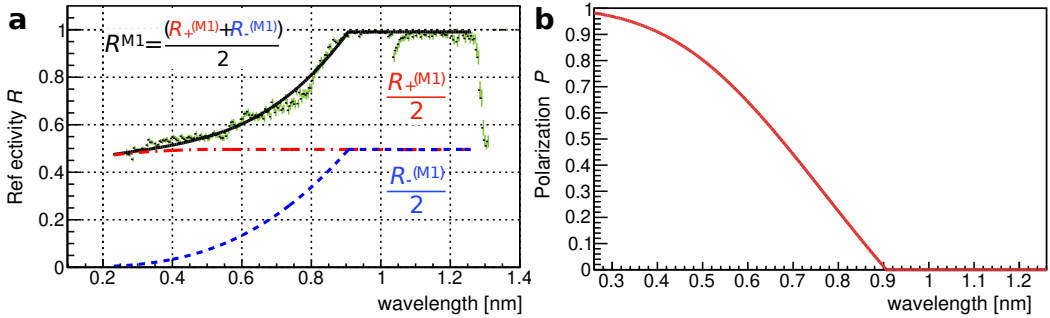}
 \centering
 \caption{\textbf{a}
 Wavelength dependence of M1 reflectivity $R^{\mathrm {(M1)}}$ (green) and reflectivity models for 
 spin $+$ state $R^{\mathrm {(M1)}_+}$ (red), 
 spin $-$ state
 $R^{\mathrm {(M1)}_-}$ (blue), which are components of $R^{\mathrm {(M1)}}$(black). 
 The black curve is a function of the M1 reflectivity fitted with a model of the reflectivity of a beam containing 50$\,\%$ each of spin $+$ state and
 spin $-$ state
 \cite{Rup}.
 The angle of M1 to the beam $\theta_1=8.3\,\mathrm {mrad}$ during the reflectivity measurement of M1 was corrected to $\theta_1=10\,\mathrm{ mrad}$ during the  reflectivity measurement of samples and the reflectivity of the mirror $R_{\rm M}$.
 \ref{fig:polarization}{b} Wavelength dependence of the polarization
 $P=(R_{+}-R_{-})/(R_{+}+R_{-})$. $P$ is obtained from the fit results of $R_{+}$ and $R_{-}$ of Fig. \ref{fig:polarization}{a}.
 }
 \label{fig:polarization}
 \vspace{-8mm}
 \end{figure}

The measurement requires a polarized beam of neutrons. To polarize an unpolarized beam from the beamline, a magnetic supermirror is utilized.
Owing to the difference of the magnetic potentials of the different spin states of neutrons, the most of reflected neutrons from the supermirror are dominated by 
neutrons of the spin $+$ state. Before measuring the actual sample, the polarized beam from the mirror was characterized with an experimental setup in Fig. \ref{fig:setup}a. 
The beam was collimated to  $\mathrm{(Vertical)}\,$10$\,\mathrm {mm}\times\,\mathrm {(Horizontal)}\,$0.1$\,\mathrm {mm}$ by a collimator C1 and then injected to a magnetic supermirror M1 which is composed of Fe, Ge and Si layers \cite{KUR}.
Both the transmitted neutrons and the reflected neutrons are detected by a two-dimensional detector based on a resistance photomultiplier tube (RPMT) \cite{RPMT}, the reflectivity of M1 is obtained from the ratio of integrated signals of transmitted and reflected neutrons. 
 Here, 
 losses due to scattering and absorption by M1 are  supposed to be negligibly small. 
 Assuming an unpolarized incident beam, the measured reflectivity of M1 $R^{\mathrm {(M1)}}$ is given by 
\begin{equation}
R^{\mathrm{(M1)}} (\lambda) = \frac{1}{2} \left( R^{\mathrm{ (M1)}}_+ (\lambda) + R^{\mathrm{ (M1)}}_- (\lambda)\right)
\end{equation}
where the M1 reflectivities of the 
spin $+$ state and 
spin $-$ state
are expressed as $R^{\mathrm{(M1)}_+}, R^{\mathrm{(M1)}_-}$, respectively.
Measured neutron intensity detected in the setup of Fig. \ref{fig:setup}a is shown in Fig. \ref{fig:setup}b. The detector signals are integrated in the ranges indicated in  Fig. \ref{fig:setup}b to obtain the intensities of the transmitted neutrons $I^{\mathrm {(M1)}}_T$ and the reflected neutrons $I^{\mathrm{(M1)}}_R$.
The reflectivity of M1 $R^{\mathrm {(M1)}}$ and the wavelength-dependent polarization $P(\lambda)$ are defined as 
$R^{\mathrm {(M1)}}={I^{\mathrm  {(M1)}}_R}/{(I^{\mathrm {(M1)}}_R+I^{\mathrm {(M1)}}_T)},\,P(\lambda)= \left(R^{\mathrm{ (M1)}}_+(\lambda)-R^{\mathrm{(M1)}}_- (\lambda)\right)\\
/\left(R^{\mathrm{ (M1)}}_+ (\lambda) +R^{\mathrm{ (M1)}}_- (\lambda)\right)$.
Assuming that $R^{\mathrm {(M1)}}_+$
is constant and $R^{\mathrm {(M1)}}_-$
obeys a known model function discussed in Ref. \cite{Rup}. The reflectivties $R^{\mathrm {(M1)}}_+$ and $R^{\mathrm {(M1)}}_- $ are obtained by fitting the model to the data as shown in Fig. \ref{fig:polarization}{a}.
As a result, 
$P(\lambda)$ is obtained as shown in Fig. \ref{fig:polarization}{b}.

\subsection{Measurement of the thin iron film samples}\label{sec:polarization}
\begin{wrapfigure}{r}{0.5\textwidth}
\vspace{-3em}
  \includegraphics[width=0.5\textwidth]
  {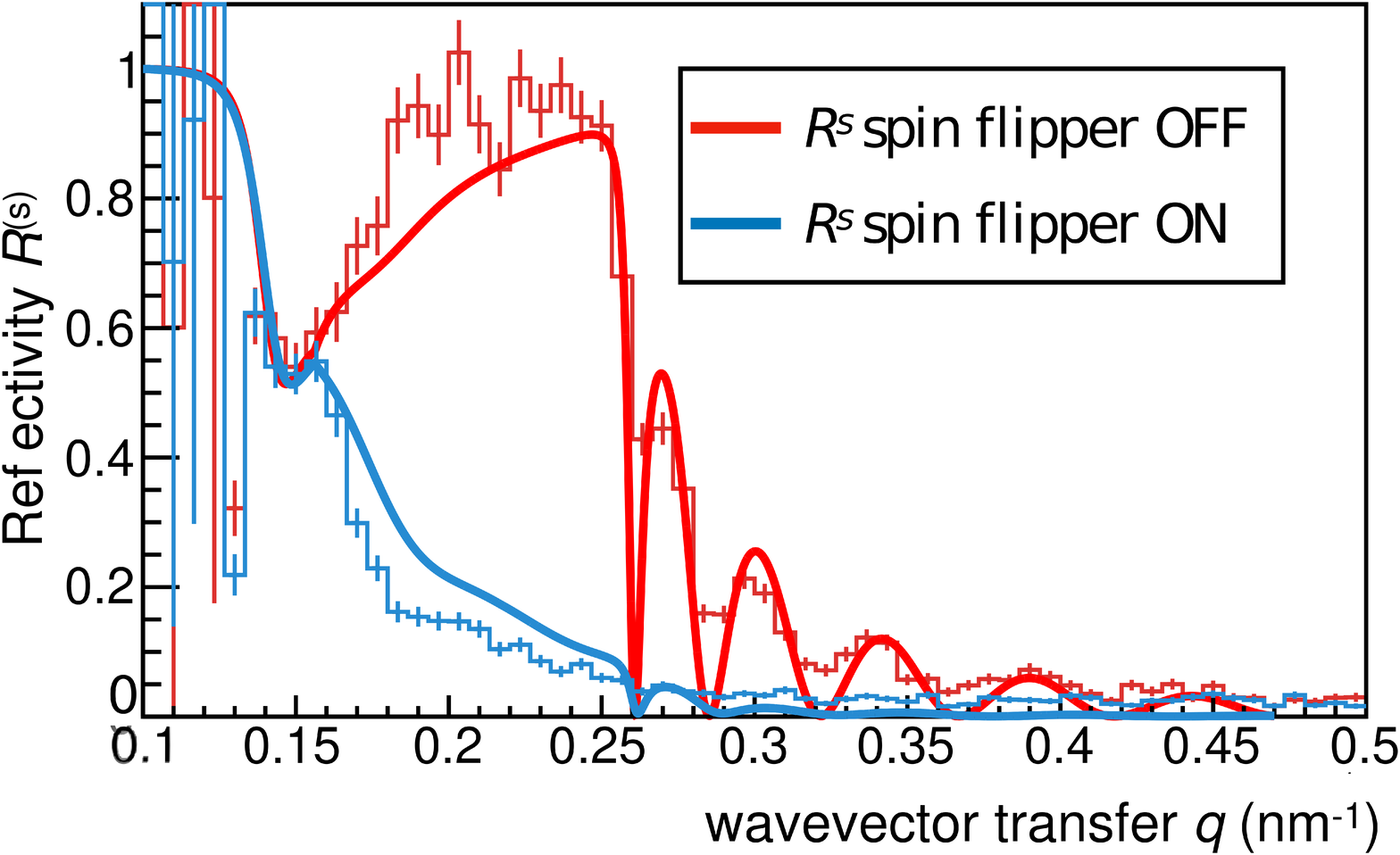}
 \centering
 \caption{ The wavevector transfer $q$ dependence of  reflectivity $R^{\mathrm {(s)}}$ of the sample with a thickness of 90$\,\mathrm{nm}$. The red histogram shows $R^{\mathrm {(s)}}$ with the spin flipper OFF and the blue histogram shows $R^{\mathrm {(s)}}$ with spin flipper ON.
 The curve is a function of the reflectivity fitted with Eq. \ref{Rs_off}.
 }
 \label{fig:fitting}
 \vspace{-2em}
 \end{wrapfigure}

The reflectivities of polarized neutrons were measured by a setup shown in Fig. \ref{fig:setup}{c} to determine the magnetic potentials of the iron film samples. As shown in Fig. \ref{fig:setup}{c}, the sample was placed in a dipole magnet with an applied magnetic field up to
$\sim $6.4$\,\mathrm{kA/m}$.
A collimator C2 $\mathrm {(Vertical)}\,
$10$\,\mathrm {mm}\times\,\\
\mathrm {(Horizontal)}\,$0.1$\,\mathrm {mm}$ was added between the M1 and  the sample  to avoid scattering of neutrons at the edge of the  sample. 
A spin-flipper based on the principle of adiabatic free passage (AFP) \cite{Grigoriev} is placed upstream of the C2 to select the spin state of incident neutrons to the sample.
Measured neutron intensities detected with this setup are shown in Fig. \ref{fig:setup}{b} shows Fig. \ref{fig:setup}{d}, where the reflectivity of samples, $R^{s}$ is defined as the ratio of direct beam intensity  $I^{\mathrm {(s)}}_D$ to reflection intensity of the sample
$I^{\mathrm {(s)}}_R$.
The measured reflectivity of the sample with a thickness of 90$\,\mathrm{nm}$, $R^{\mathrm {(s)}}$ is shown in Fig. \ref{fig:fitting}.
The $R^{\mathrm {(s)}}$ is related to the beam polarization and the reflectivity function $R(q~|V,d)$ as
\begin{equation}
R{\mathrm {(s)}}(q)= \left\{
    \begin{aligned}
    & \frac{{1+P(q)}}{2} R(q~|V_{\mathrm {eff}\,+},d)  + \frac{{1-P(q)}}{2}R(q~|V_{\mathrm {eff}\,-},d) \quad (\mathrm {when}\ \mathrm{SF=OFF})\\
    & \frac{{1+P(q)}}{2}R(q~|V_{\mathrm{ eff}\,-},d)   + \frac{{1-P(q)}}{2}R(q~|V_{\mathrm {eff}\,+},d) \quad (\mathrm {when}\ \mathrm{SF=ON})
    \label{Rs_off}
    \end{aligned}
\right.
\end{equation}
where 
$V_{\mathrm {eff}\,+},\,V_{\mathrm {eff}\,-}$
are the effective potentials experienced by the neutrons in each spin state. The polarization $P(\lambda)$ obtained in Sec. \ref{sec:polarization} was converted to the wavevector transfer $q$ by $q=4\pi \sin \theta_2/\lambda$, with $\theta_2$ being the incident angle of the sample in this setup.
Fitting the functions of Eq. (\ref{Rs_off})
to the measured data in Fig. \ref{fig:fitting}, the effective potentials are evaluated as  
$V_\mathrm{eff\,+}=328(1)\,\mathrm {neV}$
and
$V_\mathrm{eff\,-}=65(1)\,\mathrm {neV}$.
From these,  $V_{\mathrm {Fe}}=196(1)\,\mathrm{neV},\,B=2.18(1)\,\mathrm {T}$ and $\, d=94.8(4)\,\mathrm {nm}$ are determined. 

\section{Conclusions and Outlook}
The thin iron films for polarization analysis of UCNs are successfully developed and characterized.
From the results of the VSM measurements, the magnetic field required to saturate the iron films was obtained as $\sim 4 \,\mathrm{kA/m}$ for films on the Si substrates and 12$\,\mathrm{kA/m}$ for those on the Al substrates. The cold neutron reflectometry of the Si-substrate films were conducted,  and yielded  the effective potentials of 
$V_\mathrm{eff\,+}=328(1)\,\mathrm {neV}$ and 
$V_\mathrm{eff\,-}=65(1)\,\mathrm {neV}$,
corresponding to the saturation magnetization of 2~T. These characteristics satisfy the requirements for the use as a part of the TUCAN SSA. As the next step, a prototype UCN polarizer is planned to be constructed and tested by pulsed UCNs provided from J-PARC/MLF BL05 \cite{Imajo2016}.

\section{Acknowledgements}
\par
This research was supported by JSPS KAKENHI
Grant Numbers 18H05230 and 20KK0069.
The neutron experiment at the Materials and Life Science Experimental Facility of  J-PARC was performed under user S-type project of KEK (Proposal No. 2019S03).  The thin iron films production  work has been carried out under the visiting Researcher’s Program of the Institute for Integrated Radiation and Nuclear Science, Kyoto University.

\end{document}